\documentclass{article}

\usepackage[nonatbib]{neurips_2021_2}
\usepackage{graphicx,subfigure}

\usepackage{graphicx,subfigure}

\usepackage{amsmath}

\usepackage{amsmath}
\usepackage{graphicx}
\usepackage{subfigure}
\usepackage{makecell}
\usepackage[utf8]{inputenc} % allow utf-8 input
\usepackage[T1]{fontenc}    % use 8-bit T1 fonts
\usepackage{hyperref}       % hyperlinks
\usepackage{url}            % simple URL typesetting
\usepackage{booktabs}       % professional-quality tables
\usepackage{amsfonts}       % blackboard math symbols
\usepackage{nicefrac}       % compact symbols for 1/2, etc.
\usepackage{microtype}      % microtypography
\usepackage{xcolor}         % colors

\let\svthefootnote\thefootnote
\newcommand\freefootnote[1]{%
  \let\thefootnote\relax%
  \footnotetext{#1}%
  \let\thefootnote\svthefootnote%
}

\title{A Survey of Deep Learning Methods in Protein Bioinformatics and its Impact on Protein Design  }

\author{%
  Weihang Dai\thanks{Use footnote for providing further information
    about author (webpage, alternative address)---\emph{not} for acknowledging
    funding agencies.} \\
  Department of Computer Science and Engineering\\
  Hong Kong University of Science and Technology\\
  Hong Kong SAR\\
  \texttt{wdai03@gmail.com} \\
}

\begin{document}

\maketitle

\begin{abstract}
Proteins are sequences of amino acids that serve as the basic building blocks of living organisms. Despite rapidly growing databases documenting structural and functional information for various protein sequences, our understanding of proteins remains limited because of the large possible sequence space and the complex inter- and intra-molecular forces. Deep learning, which is characterized by its ability to learn relevant features directly from large datasets, has demonstrated remarkable performance in fields such as computer vision and natural language processing. It has also been increasingly applied in recent years to the data-rich domain of protein sequences with great success, most notably with Alphafold2's breakout performance in the protein structure prediction. The performance improvements achieved by deep learning unlocks new possibilities in the field of protein bioinformatics, including protein design, one of the most difficult but useful tasks. In this paper, we broadly categorize problems in protein bioinformatics into three main categories: 1) structural prediction, 2) functional prediction, and 3) protein design, and review the progress achieved from using deep learning methodologies in each of them. We expand on the main challenges of the protein design problem and highlight how advances in structural and functional prediction have directly contributed to design tasks. Finally, we conclude by identifying important topics and future research directions.  

\end{abstract}

\freefootnote{Submission: Oct 2021}

\section{Introduction}
% briefly introduce the exiting methodologies / what has been achieved so far. 
 
 Proteins form the basic building blocks of organisms and perform various functions through chemical reactions with molecules. They are composed of chains of 20 possible amino acids that stabilize to a 3D structure based on their sequence. A protein's sequence, structure, and function are closely related: sequence largely determines structure, and structure largely determines functionality. The ability to understand proteins, from their initial structural conformation to their functional properties after stabilization, are vital for bioengineering tasks such as drug design, disease identification, and therapeutic treatment. 
 
 %godzik2011metagenomics
 %has some good data and figures on size of proteins
 Improving technologies for protein sequencing and structure determination have led to the rapid growth of various databases such as PDB \cite{berman2002protein}, UniRef50 \cite{suzek2015uniref}, BindDB \cite{bader2001bind}, and many others, which collect structural and functional information of various sequences. Despite the growing knowledge of existing sequences and their properties, our understanding of the natural processes that drive them remain limited. The inter- and intra-molecular forces involved are highly complex, making traditional statistical and physical modelling approaches challenging. The possible orderings and combinations of amino acids also form a massive sequence space \cite{godzik2011metagenomics}, and existing databases are only limited to the space explored by natural evolution \cite{huang2016coming}. Important tasks such as \textit{de novo} design however, which involves designing proteins rarely occurring in nature, require an intimate understanding of the relationship between sequence, structure, and function.  
 
 Deep learning techniques have been increasingly applied to the field of bioinformatics in recent years. These techniques are characterized by their ability to learn relevant features automatically through gradient descent using large datasets, and have been shown to consistently outperfrom traditional modelling techniques in different fields. The increasing availability of protein sequencing data has made protein bioinformatics a particularly suitable domain for deep learning. This is particularly apparent in the protein structure prediction problem, which aims to determining the 3D structure of a protein when only given its sequence. In 2020, Deep Mind's Alphafold2 model was able to achieve structural prediction accuracy comparable to experimental validation methods in the CASP14 competition, arguably "solving" the problem. Impressive progress has also been made in other problems such as protein functional prediction, protein-protein interaction (PPI), binding site identification, \textit{etc} by similarly applying deep learning techniques. 
 
 Protein design remains a challenging problem however since it involves by definition discovering sequences or structures that may not occur naturally. Whereas tasks such as structural and functional prediction are relatively well defined and accompanied by relevant datasets, protein design requires discovering stable molecules subject to some functional objective. This also happens to be the most relevant for problems such as drug design and disease treatment and have the largest potential for social benefit. 
 
 In this paper, we systematically review the progress that has been achieved from applying deep learning methodologies to protein bioinformatics and how they have unlocked new possibilities for improvements in protein design. 
%  In this review, we systematically explore the main topics related to protein bioinformatics and survey the progress that has been achieved from applying deep learning methodologies. The study of proteins largely involves understanding the relationship between protein sequence, structure, and function. Consequently, w
 We follow the common paradigm of broadly classifying problems into three main categories \cite{dill2017protein}: 1) structural prediction, 2) functional prediction, and 3) protein design, and start by first reviewing recent progress for categories 1) and 2), which are well defined problems addressable with large labelled datasets. We then discuss the main challenges of 3), recent progress from applying deep learning techniques, and how breakthroughs in 1) and 2) have also led to improvements in design by treating it as the inverse of structural and functional prediction. Finally, we conclude by discussing potential future research directions in applying deep learning to protein design.

\section{Background}
\subsection{Key concepts of deep learning}

Deep learning has seen rapid advancements in recent years and achieved remarkable performance on a wide variety of computer vision (CV) and natural language processing (NLP) tasks. These advancements have also been successfully applied to other fields, including bioinformatics. In this section, we briefly cover some of the key ideas and advantages of deep learning compared to traditional modelling. We then review some of the common architectures used that are also relevant to this survey.

\subsubsection{Learning through gradient descent}

Deep learning architectures typically refer to models consisting of multiple layers of stacked neural networks. One of the simplest neural network architectures is the single layer perceptron (SLP). For some model $f(\cdot)$, the single layer perceptron is defined by:

\begin{equation}
    f(x) = \alpha(w^T x + b)
\end{equation}

where $w$ is a weight vector, $b$ is a bias term, and $\alpha$ is a non-linear activation function such as sigmoid, tanh, or ReLU. Neural networks are trained through back-propagation, which involves updating weight parameters iteratively based on a pre-determined loss function. For some dataset $\{x_i, y_i\}^N_{i=1}$ where $X$ is the sample input, $Y$ is the sample label, and $N$ is the total number of samples, we obtain our label estimate using our SLP, $f(x_i) = \hat{y_i}$, and calculate our loss using some pre-determined loss function, $\mathcal{L}(\hat{y}, y)$. Our goal is to find the optimal parameters for $w,b$ that minimizes the loss, \textit{i.e.} 

\begin{equation}
    \min_{w,b} \mathcal{L}(f(x;w,b), y) .
\end{equation}

\begin{figure}
\centering     %%% not \center
\subfigure[Figure A]{\label{fig:a}\includegraphics[width=65mm]{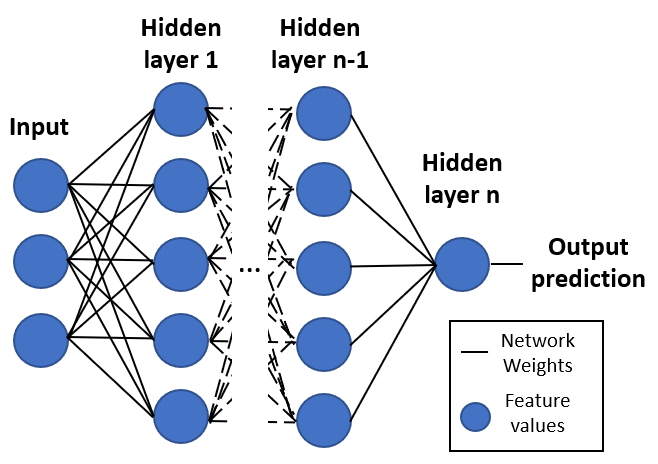}}
\subfigure[Figure B]{\label{fig:b}\includegraphics[width=65mm]{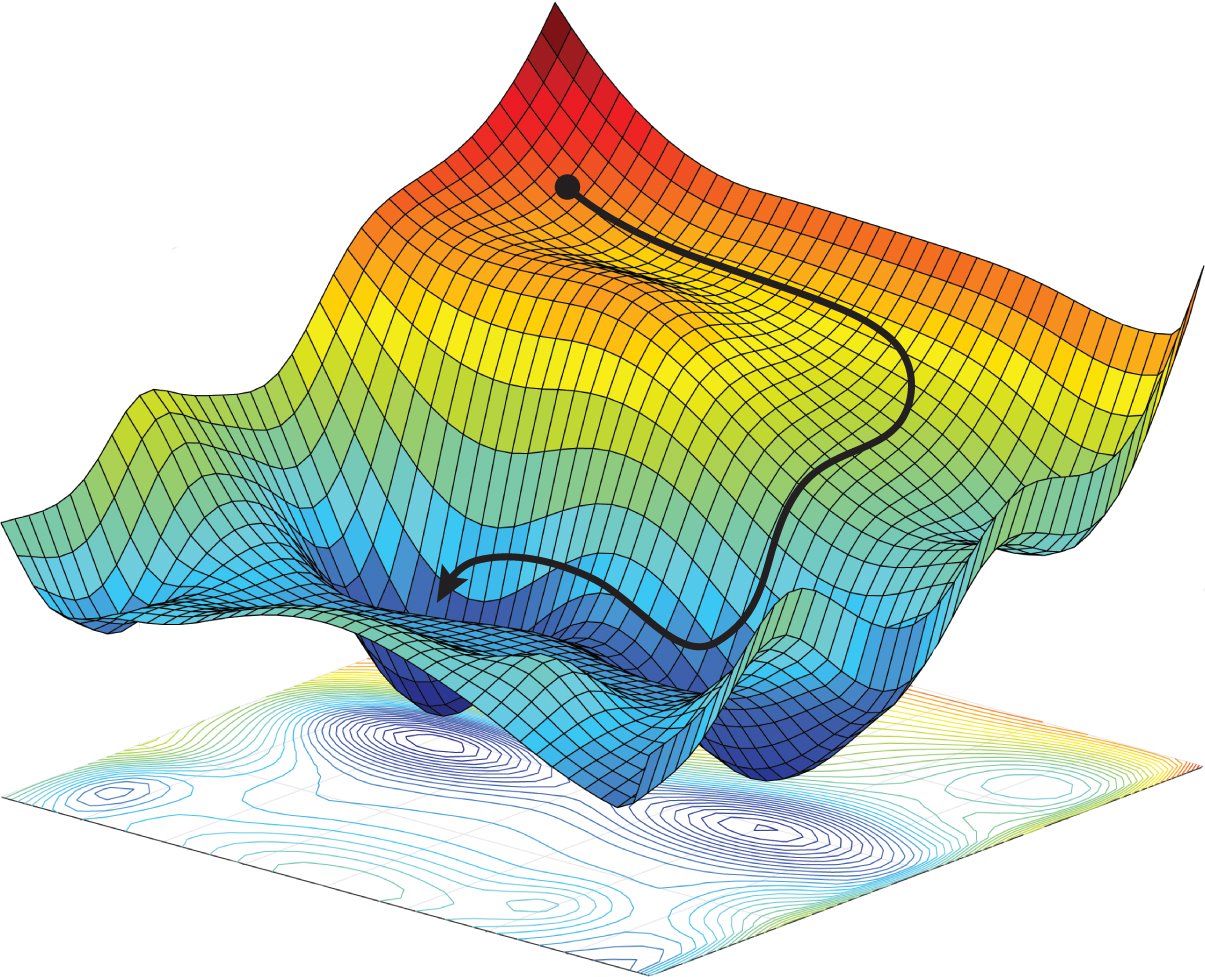}}
\caption{(a) Simplified representation of a MLP with n layers. Input and intermediate features are passed on along the network through matrix operations and non-linear activation functions. (b) A simplified 3D representation of stochastic gradient descent. Weights are updated such that the final values correspond to a minimum on the loss surface. Figure taken from \cite{amini2018spatial}}
\end{figure}

The minimization process is done through stochastic gradient descent, which involves a forward and backward propagation step. During each iteration time-step, $t$, a set of inputs are fed into the network to obtain a prediction and loss value through forward propagation. Backward propagation is used to calculate the gradient of the parameters with respect to the loss value and update weights based on the gradient. We can describe this process as follows:

\begin{equation}
    w_{t+1} = w_{t} - \gamma \frac{L(f(x;w_t,b_t), y)}{w_t}
\end{equation}

where $\gamma$ is the learning rate parameter. This process can be seen as navigating the loss surface to some minimum parameter configuration, as represented in Figure \ref{fig:b}.The SLP can be made deeper by stacking numerous layers on top of each other, forming a multi-layer perceptron (MLP), which is described in Figure \ref{fig:a}. MLPs typically do not perform well and have a tendency to over-fit to the training data. This is because no additional constraints are imposed on the large number of parameters and the model fails to learn relevant representations. To correct this, concepts such as spatial invariance and equivariance have be introduced into network operations. Examples of these are convolutional operations for translational equivariance in Convolutional Nueral Networks (CNNs) and translational invariance using pooling operations. These effects combine to reduce model search space. For a more detailed and technical explanation, we refer readers to \cite{goodfellow2016deep}. 

\subsubsection{Comparison with traditional modelling approaches}

One of the key advantages of deep learning over traditional modelling approaches is that deep learning does not rely on manually determined features. Instead, the model is able to learn a large set of relevant features automatically during the training process. These features have been demonstrated to perform better than manually defined ones, since manually features are often determined based on fragile, biased, or incorrect assumptions. 

An example of this is the set of manually defined filters in SIFT \cite{lowe1999object} for CV applications. Experiments have shown that deep learning architectures are able to learn similar filters in the different layers of their convolution kernels, but in much larger numbers \cite{simonyan2014very}. Because these are learnt automatically during the training process, they also tend to be more accurate and relevant to the task. 

Another key advantage is that neural networks are universal functional approximators \cite{hornik1989multilayer}, which is particularly useful when modelling complicated systems that are difficult to explicitly define. Universal approximation only guarantees convergence on a given dataset however and does not ensure correct causal relationships are learnt. For this, other constraints and experimental methodologies are required, such as performance verification on separate validation and test sets. 

The disadvantage of deep learning is that the amount of data and computation requirement is significantly higher compared to traditional modelling. However, given the increasing availability of computation power and large datasets, this is becoming less of an obstacle. Advancements in fields such as unsupervised and semi-supervised learning have also allowed us to train accurate models even in the presence of limited labelled data \cite{van2020survey, bengio2012deep}. 

\subsubsection{Common Deep Learning Architectures}

\textbf{Convolutional Neural Networks (CNNs) - } CNNs are a class of deep learning models that use the convolutional neural networks as core components of its model. The convolution kernels serve as filters with weights that resemble manually defined kernel filters after training. The use of convolutional filters also reduce the number of model parameters through weight sharing, thus enabling translational equivariance. This class of models include VGG \cite{simonyan2014very}, ResNets \cite{he2016deep}, DenseNets \cite{huang2017densely}, and others. 

\textbf{Recurrent Neural Networks (RNNs) - } RNNs capture sequential information by forming connections between nodes in the temporal domain. They often contain an internal memory state that allows it to retain information from previous time-steps when processing new time-steps. RNNs are typically used for sequential inputs in NLP problems although they have also been adapted to CV tasks. 

\textbf{Auto-encoders (AEs) / Variational Auto-encoders (VAEs) - } AEs/VAEs are networks trained with the specialized task of learning a representational coding of data. They typically consist of an encoder and decoder, and are characterized by a bottle-neck structure, where the data is reduced in dimension in the middle. The bottle-neck structure effectively forces the network to learn an efficient low-dimensional coding representation for the data. The network is usually trained with reconstruction loss in a unsupervised setting.  

\textbf{Generative Adversarial Networks (GANs) - } GANs are networks trained for generative tasks. They are trained with both a generator network, trained to generate the target outputs, and a discriminator network, trained to determine whether a sample is real or generated. Training with two networks allows the output to better reflect the training distribution.

\textbf{Transformers - } Transformers were first used in NLP problems and were shown to perform better than RNNs, which are vulnerable to exploding gradients during training. These architectures use multi-head self attention modules that allow the network to learn which inputs and features to focus on across a larger input window \cite{vaswani2017attention}. Transformers were later shown in \cite{dosovitskiy2020image} to outperform traditional CNNs for computer vision problems as well. The major advantage of transformers over traditional CNNs is that transformers are able to consider long-range relationships between inputs, whereas CNNs are limited by the field of view of its kernels. Because of its more general form however, the original transformer architecture required large amounts of data and are difficult to train. Subsequent advancements such as the SWIN transformer \cite{liu2021swin} and Reformer \cite{kitaev2020reformer} have made improvements in this area however. Transformers are playing an increasing role in various domains, including bioinformatics, and are increasingly adopted in model architecture designs. 

\subsubsection{Architectures for Geometric Deep Learning}

Architectures mentioned in Section 2.1.3 are typically used for problems in the temporal space, 2D Euclidean plane, 3D Euclidean space. Many important problems are outside of these standard spaces however and may involve graphs, groups, or manifolds. This is particularly true for proteins as it is difficult to directly process protein structures as 3D volumes. Geometric Deep Learning is a more generalized framework that covers feature transformation and aggregation operations for a variety of input domains, including graphs, manifolds, groups, and also the standard Euclidean spaces. We give a brief overview of the network classes relevant to protein bioinformatic problems. For a more detailed treatment, we refer readers to \cite{bronstein2021geometric}.

\begin{figure*}%
\centering
\includegraphics[width=1 \columnwidth]{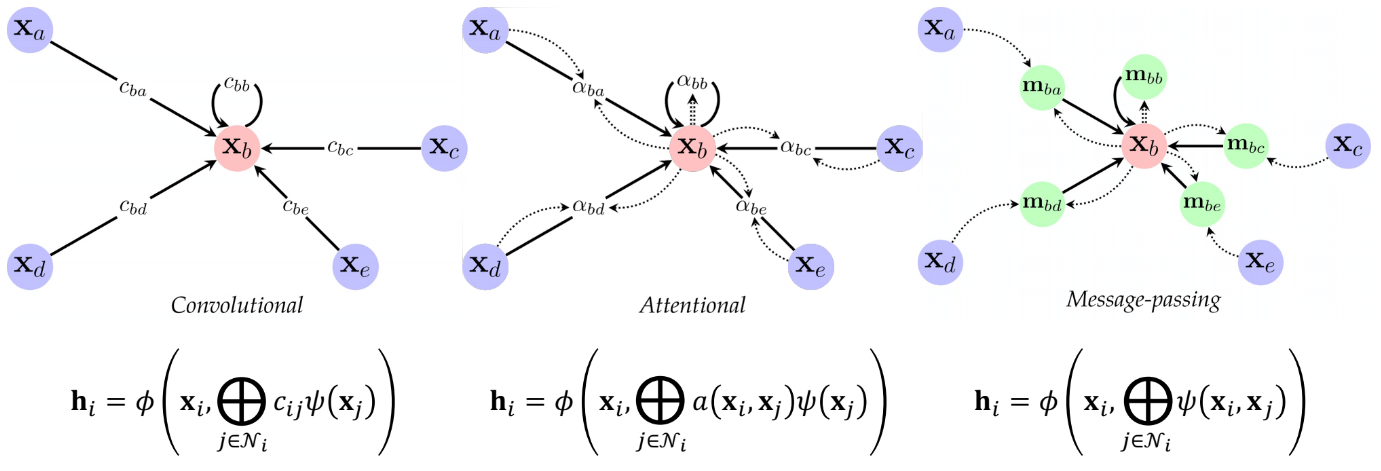}%
\caption{A simplified overview of the main classes of GNNs. GNNs all involve a convolutional operator, $\psi$, a pooling operator, $\bigoplus$, and a non-linear activation function, $\phi$. Variations between them can be classified depending on the weighting methodology for features from neighbouring nodes. Figure taken from \cite{bronstein2021geometric} }
\label{GNN_overview}
\end{figure*}

\textbf{Graph Convolutional Neural Networks (GCNNs) - } GCNNs operate on graph data, where sample inputs consists of vertices and edges between them. Features are represented as vertices, and edges are used to determine feature aggregation between vertices during learning. Graphs are a highly general representation for different data types, and can also be used to represent 1D, 2D, and 3D Euclidean data by treating the inputs as a grid. Equivalents of convolution, pooling, and attention operators in Euclidean space data are also used for feature extraction for GNNs. Figure \ref{GNN_overview} gives a generalized overview of different GNN structures. 

\begin{figure*}%
\centering
\includegraphics[width=0.8 \columnwidth]{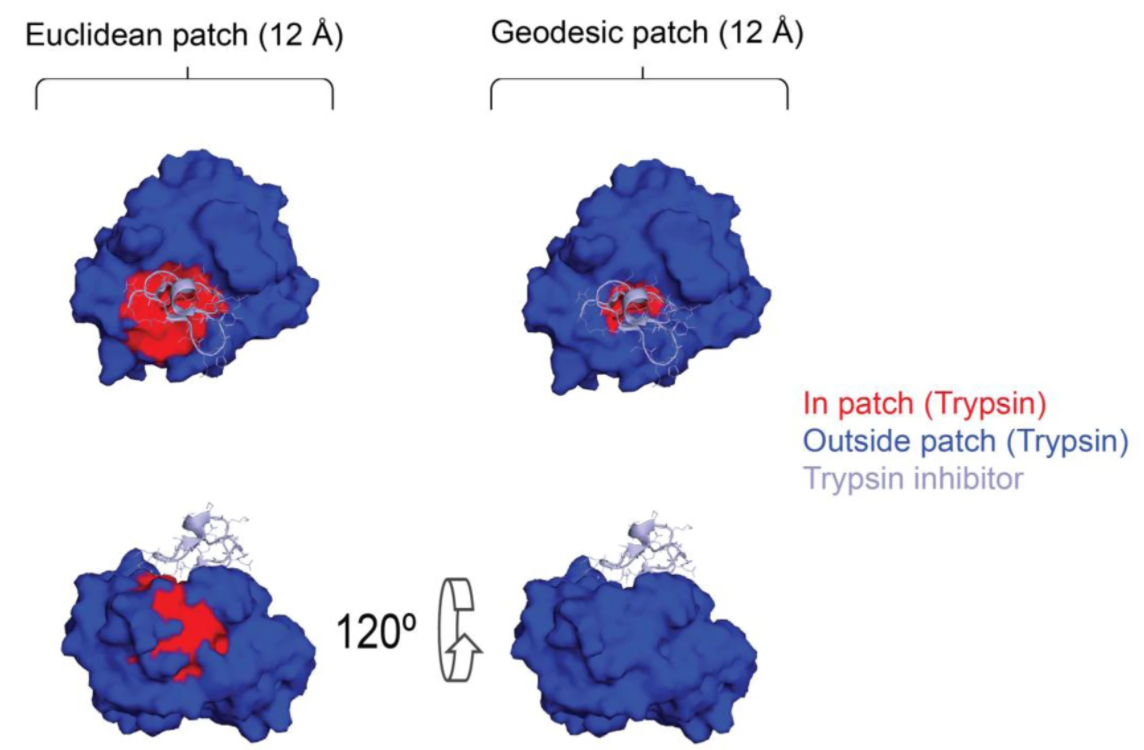}%
\caption{An Euclidian patch covers the region that is enclosed by a circular projection from a 2D Euclidian plane tangent to the surface. A geodesic patch covers the 2D region that is within some distance from a central point when measured along the surface. The example shows differences between the two patches on a protein structure with a deep pocket in the binding site. In the bottom left, the Euclidian patch covers a large region away from the tangent point that is irrelevant to the binding site. The Geodesic patch only covers the surface that is part of the binding site. Figure taken from \cite{gainza2020deciphering}}
\label{geodesic_patch_enzyme}
\end{figure*}

\textbf{Geodesic CNNs - } Geodesic CNNs are used on manifolds and intuitively adapt the convolutional operator to irregular surfaces. Geodesic patches are patches on a manifold where the edges are of the same length to a center-point. Different feature aggregation and pooling functions can then be applied on the patches for feature learning. Extracted features can then be used with different predictive heads for different tasks similar to standard deep learning architectures. Figure \ref{geodesic_patch_enzyme} illustrates the difference between standard Euclidean and geodesic metrics on surfaces. 

\subsection{Protein Bioinformatics}

Bioinformatics is the application of computer science methodologies to the treatment of biological data, mainly DNA and proteins, although we limit our scope to proteins in this paper. We provide a brief background to proteins and give an overview of the main problem classes in protein bioinformatics.

\subsubsection{Protein composition and interaction}
\begin{figure*}%
\centering
\includegraphics[width=0.85 \columnwidth]{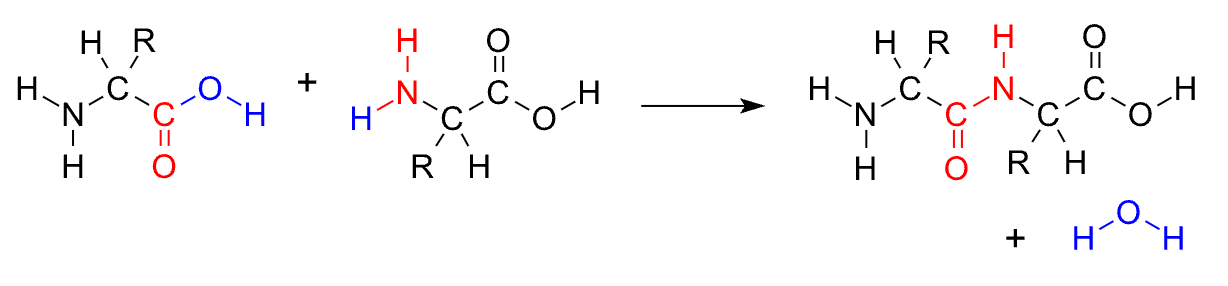}%
\caption{ Peptide bonds are formed between the Carbon and Nitrogen atoms in two amino-acids, releasing water in the process. The backbone of a protein structure is composed of these Carbon and Nitrogen atoms and is typically described by the 3D configuration of $C_{\alpha}$ atoms. }
\label{chem_react}
\end{figure*}

Proteins are chains of amino acids joined together by peptide bonds. The bond joins the carbon atom of one amino acid chain with the nitrogen atom on another amino acid chain, releasing a water compound in the process. An example of this is shown in Figure \ref{chem_react}. The amino acid residue forms as a side-chain on what is referred to as the $C_{\alpha}$ atom. Consequently, the backbone of an amino acid chain can be described by a series $C_{\alpha}$ atoms in 3D space. The side-chain residues determine the various properties of the protein molecule and also restrict the structural conformation.

% Figure taken from \cite{wikipedia_2021}

Protein structures are generally described using the term primary, secondary, and tertiary structures. Primary structure refers to the generated but unfolded sequence. Secondary structures are those formed when folding first begins and consists of alpha-helices or beta-sheets. This process happens quickly and is driven by hydrogen bonds within the backbone chain. Tertiary structures are formed through interactions between the secondary structures and determined by surface hyrdophobic/hydrophilic properties as well as other forces. Bonds between residue pairs may also form. Sometimes, the formed structures also interact with additional amino-acid chains to form what are known as quaternary structures.

\begin{figure*}%
\centering
\includegraphics[width=0.85 \columnwidth]{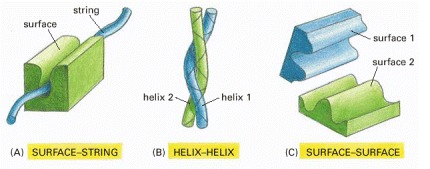}%
\caption{Simplified examples different kinds of protein binding. Surface-surface binding is the most common type. Figure taken from \cite{alberts_1970} }
\label{bind_graph}
\end{figure*}

Protein folding is a dynamic process that releases energy and are driven by hydrophilic/hydrophobic forces, Van der Walls forces, and conformational entropy \cite{anfinsen1973principles, levinthal1968there}. In most cases, the structures stabilize at minimum free entropy, although it is possible that they stabilize at a higher energy level because they are unable to dynamically reach a lower configuration. There are sometimes variations in the folds, and it is possible that some sequences stabilize to different structures depending on their external environment. On the other end of the spectrum, proteins with the same structures sometimes may have slightly different sequences due to evolutionary mutation between and within species. 

%https://www.ncbi.nlm.nih.gov/books/NBK26911/ \cite{roberts2002molecular}
Proteins perform functions through chemical reactions with surrounding molecules. More specifically, proteins bind to highly specific targets, which are referred to as ligands, through non-covalent bonds, bonds where electrons are not shared between atoms, on their surfaces \cite{alberts_1970}. Binding sites of proteins fit tightly with target ligands like a glove, as shown in Figure \ref{bind_graph}. Individual non-covalent bonds, which are determined by properties of surface poly-peptide residues, are extremely weak, but the tight fit combines many individual bonds to strengthen the binding force. In some cases, specific characteristics of the protein surfaces contribute to the binding strength. For example, in some structures there may exist clusters of negatively charged residues that normally repel each other, but are held together by their structure to attract positive ions. 
Figure \ref{bind_graph} shows some of the typical types of protein binds.

\subsubsection{Key problems in protein bioinformatics}

Protein properties are primarily determined through manual experimental approaches. This means our knowledge of the universe of proteins are limited to those that are naturally existing and bottle-necked by the speed of experimentation. Because of this, many bioengineering problems, especially the task of \textit{de novo} protein design, remain particularly challenging. A large part of protein bioinformatics revolves around discovering relationships between protein sequence, structure, and function, using computational methods. Based on this, we broadly classify the problems into 3 categories:

 \begin{enumerate}
     \item Structural prediction from sequence \\
        The problem of determining the structure of protein purely based on its sequence (sequence $\rightarrow$ structure). This is generally a well defined problem with public challenges that are held regularly to benchmark process. 

     \item Functional prediction from sequence or structure \\
        The problem of predicting functional properties of a protein given its sequence or structure (sequence $\rightarrow$ function || structure $\rightarrow$ function). This includes tasks such as functional prediction, binding site prediction, protein classification, PPI \textit{etc}. and is much more varied
        
    \item Protein design \\
        The problem of determining stable sequences given some desired functional property or pre-determined structure (function $\rightarrow$ sequence || structure $\rightarrow$ sequence). These tasks are the most difficult as multiple sequences may conform to the same structure or perform the same function. Generated proteins must also be synthesized experimentally for truly accurate validation, thus limiting the availability to evaluate and benchmark performance
 \end{enumerate}
 
\begin{figure*}%
\centering
\includegraphics[width=0.85 \columnwidth]{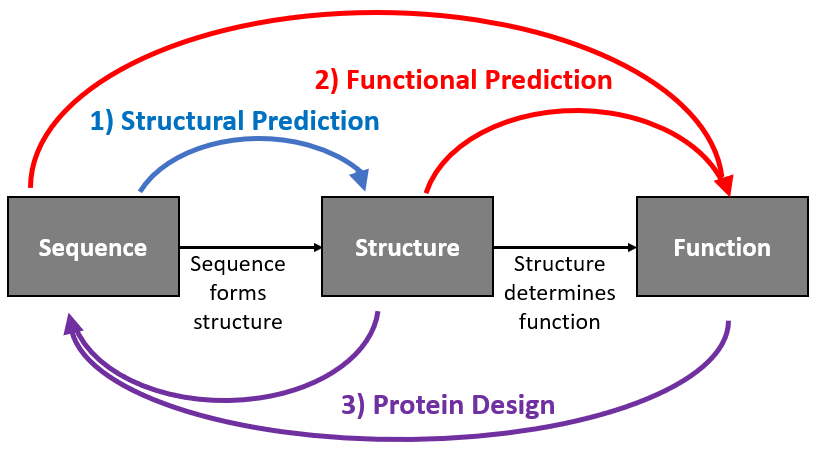}%
\caption{Visual relationship between the three main categories of problems in protein bioinformatics. Protein design can be seen as the inverse of structural and functional prediction, and some methodologies directly follow this intuition. }
\label{problem_class}
\end{figure*}

 Figure \ref{problem_class} shows the general relationship between these 3 classes of problems.

\section{Structural prediction from sequence}

Protein structures are currently determined through techniques such as NMR spectroscopy and X-ray crystallography. The problem of protein structure prediction from amino-acid sequences involves inferring structure computationally from sequences without the need for manual experimentation. Known structural information on tested proteins sequences are stored in large data banks such as the PDB, which contains information on over 180,000 sequences. The problem is therefore well defined with readily available datasets. 

Public competitions as such as CASP and CAMEO are held in regular intervals to benchmark progress so far in structural prediction. Although neural networks have been used for this problem from as early as 1997 \cite{lund1997protein}, the competition was dominated by traditional statistical modelling for many years. Deep learning approaches only became a serious alternative in CASP12 with the appearance of Raptor X \cite{wang2017accurate}. This was followed by Deep Mind's win in CASP13 with Alphafold1 \cite{senior2020improved} and their incredible performance in CASP14 with Alphafold2 \cite{jumper2021highly}, which achieved near experimental accuracy. 

In this section, we first briefly go over key ideas in structural prediction in traditional modelling approaches. We then review the common deep learning techniques that have been used so far, as well as the current state-of-the-art approach.

\subsection{Traditional methods and key ideas}

Traditional approaches to protein fold modelling are based on searching through the space of possible structural configurations to find the minimum entropy configuration. The search process is computationally expensive however, although a number of key ideas have been utilized to reduce the search space and improve accuracy.

\textbf{Homogolous search and threading - } The simplest approach to structure prediction is to approximate the structure with highly similar sequences or sub-sequences with known structures. This can be effective if close sequences exist, but have very limited performance when this is not the case. Nevertheless, the tools developed through this approach are still widely for finding closely aligned sequences of proteins, which are referred to as multiple sequence alignments (MSAs). 

Sequence alignment through large protein databases is an algorithmic challenge. The most successful algorithm, still often used to this day, is the PSI-BLAST algorithm, proposed in \cite{altschul1997gapped}. PSI-BLAST builds upon the BLAST algorithm, which performs alignment search by optimizing for a similarity heuristic instead of performing identity matching between two sequences, thus improving efficiency and accounting for mutations. After the BLAST algorithm is run to find alignments with high similarity scores, alignments are used to calculate a position specific scoring matrix (PSSM), which reflects the probability of a amino acid appearing at a sequence position. The PSI-BLAST algorithm then uses the PSSM to improve further alignment searches whilst refining the PSSM value iteratively. PSSMs capture important information within MSAs are also frequently used as inputs features for deep learning approaches. 

% Threading is similar in concept to homogolous search, except it allows for larger gaps and insertions between two alignments at the expense of an imposed penalty. The intuition is that by allowing for more flexible sequence alignments, subject to some difference penalty, a wider search space is considered. The THREADER algorithm was originally proposed in \cite{jones1992new}, where the authors demonstrate that it could accurately predict the structure of C-phycocyanin, which a structure similar to globin folds, but with only 14 out of 174 identity matches with the closest myo-globin protein. 

\textbf{Contact pair predictions - } Contact pairs within a protein structure refer to points within the protein backbone that are close together in the final fold. These are typically defined as pairs of $C_\beta$ atoms that are less than 8 angstroms (\r{A}; 1\r{A} = $10^{-10}m$) in distance. It has been recognized early on that accurately determining contact pairs within a protein sequence significantly helps constrain the structure search space, as it serves to fix a number of key positions in the backbone. Consequently, early works have attempted to directly predict contact points directly from sequence. Altschul in 1997 \cite{altschul1997gapped} explored using neural networks through SLPs for contact prediction. Other works such as \cite{fariselli1999neural} and \cite{hamilton2004protein} include additional features to improve accuracy for contact pair prediction to the level of 21\%, which was state-of-the-art for their time. Stacked layers in the form of MLPs were also explored in their experiments but did not perform well due to lack of data at the time. The low overall accuracy limited the usefulness of these methods.

\textbf{Evolutionary data - } One of the major insights behind protein structures is that some sequences, although vastly different, perform the same function, and consequently have similar structures. Therefore, the correlation between them contains useful information on what parts of the sequence determine their structure. This is known as co-evolution information and is obtained through what are termed evolutionary coupling analysis (ECA) techniques. \cite{gobel1994correlated} uses mutual information (MI) between two sequences of the same functional family to predict contact points. MI only captures correlation between local segments in the sequence however, and is not effective for long range relationships in the sequence, a common theme in the challenge of structural prediction. \cite{marks2011protein} improved upon this approach by introducing a measure termed "Direct Information", which includes more global information to refine contact pair prediction. Structural search based on these constraints were able obtain predictions that were within 2.7-4.8 \r{A} of the ground truth structure in their test sample of proteins. The PSICOV model in \cite{jones2012psicov}, which is still used at times as feature inputs for deep learning models, captures ECA information through computing a precision matrix, which is the inverse of the correlation matrix. Sample correlation matrices are not always singular, and their algorithm calculates a sparse inverse covariance matrix through the graphical Lasso algorithm.

\textbf{Structural Search - } The techniques mentioned above rely on the availability of known sequence-structure pairs for reference. This is not possible for newly discovered protein families, which require \textit{ab initio} folding: folding directly from energy minimization. Even after constraints are found through the above methods, a search needs to be performed through structural space to determine the minimum entropy structure. The most successful algorithm for structural search is the Rosetta algorithm \cite{rohl2004protein}, which is based on the intuition that sort fragments of residues tend to arrange themselves in similar ways. The algorithm first breaks the target sequence into sliding windows of 3 or 9 residues and uses the PSI-BLAST algorithm to find a number of top matches and their torsion angles. Monte Carlo simulated annealing search is then performed from to optimize the potential energy surface (PES), which is approximated statistically based on their frequency and locations in known structures. Prior to CASP13, Rosetta was the most popular and effective method for performing structural search after constraints were identified. The original paper cites accuracies of 3-6 \r{A} for sequences of 60 or more residues without known structural templates. 

\textbf{Molecular dynamic simulations -} We make a brief mention of molecular simulation techniques. This method attempts to use physics-based simulations to model the formulation of individual amino acids \cite{marx2009ab}, although the large computational requirements limits it mostly to proteins of small sizes. 

\textbf{Overall framework - } From the above, we can see the structure prediction process can be largely divided into a two stage process: constraint identification and structural search. Prior to Deep Mind's success in CASP, the most effective prediction algorithm was by "Zhang and QUARK" \cite{zhang2018template} which used neural networks and model ensembles to perform contact prediction before performing structure search. Common themes remain however in terms of the main difficulties, namely capturing long distance relationships within sequences, efficient structural search, and modelling of proteins with no close templates.

\subsection{Treating distance maps as a picture}

\textbf{Raptor X -} Raptor X was one of the first deep learning models that challenged traditional statistical methods in terms of performance, appearing in CASP12 \cite{wang2017accurate}. It's key innovation is that it treated the pair-wise contact map as an image, with the task of identifying residue pairs that are in contact. This allowed recent developments in the computer vision domain, such as the 2D ResNets architecture, to be carried over. The deep nature of ResNet networks allows the model to increase its focal view for lower layers, which addressed the challenge of long-range sequence relationships. Inputs consisted of the raw sequence, MSA information from PSSM and ECA infomation from the CCMpred model amongst others, whilst the output was a binary prediction of contact between a given residue pairs. The contact pair predictions were then used as constraints for structural search using the CNS tool-suite. 

\begin{figure*}%
\centering
\includegraphics[width=0.85 \columnwidth]{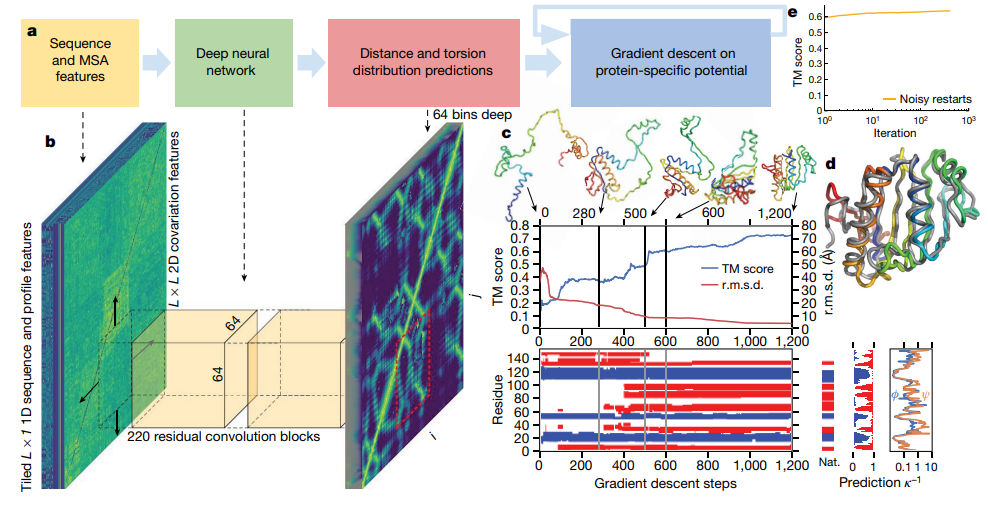}%
\caption{Network diagram of Alphafold1, taken from \cite{senior2020improved}. Similar to \cite{wang2017accurate}, their approach treats the problem of contact point prediction as a pixel classification problem, where the input matrix consists of raw sequence and MSA features. }
\label{alpha1}
\end{figure*}

\textbf{Alphafold1 -} Alphafold1 was conceptually similar to Raptor X. One of the main differences is instead binary prediction of contact, it predicted a distance distribution between pairs, which they found to improve performance significantly. Another innovation was that instead of using simulated annealing search methods, Alphafold1 was able to express the structure entirely as a series of torsion angles. The final structure was therefore fully differentiable and could be learnt through gradient descent. The flow diagram for Alphafold1 is shown in Figure \ref{alpha1}.

\textbf{trRosetta -} Rapid improvements were made upon Alphafold1's methodology after its release. Yang et al.'s trRosetta model \cite{yang2020improved} introduced additional constraints into the model by outputting orientation angles between residues in addition to residual pair distances. The residual distance, 3 dihedral angles, and 2 planar angles predicted by the model are able to fully specify the orientation between residue pairs in 3D space. Rosetta algorithms were then used to perform search based on these constraints. The additional constraints allowed them to obtain higher accuracy on targets, especially those without reference templates, compared to Alphafold1. 

\subsection{Transformers and graphs}
%https://www.blopig.com/blog/2021/07/alphafold-2-is-here-whats-behind-the-structure-prediction-miracle/
\textbf{Alphafold2 -} Alphafold2's approach in CASP14 broke away from the traditional two-stage paradigm of constraint prediction followed by structural search. Instead, an end to end model is trained, such that given an input sequence, a structural prediction is directly outputted by using a structure module, from which it is iteratively refined. Another major innovation is that instead of using MSAs directly as inputs for contact prediction, MSAs are used together with 2D residue distance maps to learn cross-attention between the mediums. 

\begin{figure*}%
\centering
\includegraphics[width=1 \columnwidth]{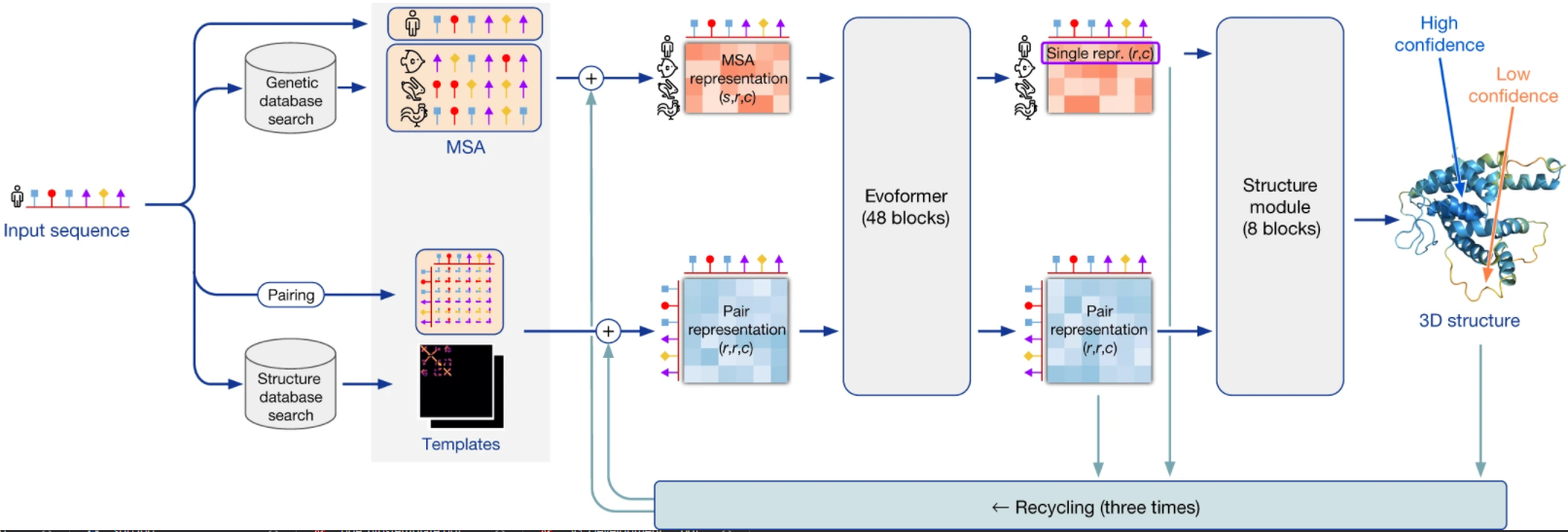}%
\caption{Alphafold2 uses a novel transformer module, the Evoformer, that allows the network to learn through self-attention the important features within sequences and contact distance maps. A structure module is also used to align individual amino acids in 3D space. Figure taken from \cite{jumper2021highly}}
\label{alpha2}
\end{figure*}

The model diagram is shown in Figure \ref{alpha2} and can be roughly divided into 3 components \cite{jumper2021highly}. In the first component, MSAs are first found using standard bioinformatic tools from protein and meta-genomic databases. This produces a number of close alignments with the target sequence from different proteins and species. Close template matches are found from the PDB structural database using HHSearch, from which pairwise residue distance maps are calclated. The second component is composed of a series of transformer modules, termed Evoformer blocks, that learn self-attention and cross-attention between MSAs and residue maps. Intuitively, correlations inferred from the MSAs can be used to guide learning from the structural templates, whilst correlations between templates can also teach the model where to focus on for MSA inputs. The third component uses the learned features from the Evoformer blocks to form structure prediction. Amino-acid residues are treated as gas particles in 3D space starting at origin. At every iteration, the structure module outputs an affine translation of the residues in 3D space until they are stable. A novel attention mechanism, termed "Invariant Point Attention" (IPA) is used to introduce invariance to translations and rotations of the overall structure in 3D space. The supplementary methodologies section of the paper contains many additional details and techniques such as the use of multiple loss functions, unsupervised learning through MSA masking and self-distillation. We only provide a review of the key concepts in this paper. 

\textbf{RoseTTAFold - } RoseTTAFold was a large community effort at replicating the results of Alphafold2 out of fear that Deep Mind would refuse to publicize their methodology. A lot of its methodology was based on a single network diagram released by Deep Mind after CASP14. A series of transformers for 1D MSA inputs, 2D template inputs, and 3D graph inputs were used to output a final model prediction, which is then iteratively refined. It uses the SE-3 Transformer, which is a self-attention module for 3D point clouds and graphs with rotation and translation equivariance constraints, as a core module. Notably, it is achieved an average TM-score of 80 on CASP14 targets versus 90 of Alphafold2, but achieved with significantly less computational resources. 

\textbf{MSA Transformer - } Transformers have also been explored for MSA data in \cite{rao2021msa} to learn unsupervised protein language models. Similar to grammar rules in everyday language, protein sequences are constrained through natural evolution to perform specific functions, thus imposing structural restraints. The model is trained similar to BERT and GPT-2 language models through the masked language modeling objective, where the model predicts the original values of masked out amino acid residues from input sequences. The attention mechanism is designed such that the model learns which sequences and where to focus on for relevant information. The MSA Transformer, designed with 100M parameters, was trained unsupervised on 26 million MSAs, each with an average of 1192 sequences. The features from the resulting model was able to outperform trRosetta in predicting contact map prediction, which also directly leads to more accurate structural models. They also found that the attention mechanism learnt by the Transformer corresponded well with contact points within the sequence, and that accurate inference is achieved even for sequences with little few alignments in existing databases. 

\subsection{Others architectures}

\textbf{Recurrent geometric networks (RGNs) } The RGN model is another approach to learning protein structure end-to-end from input sequences, such that 3D coordinates for individual $C_\alpha$ atoms are directly outputted. The model utilizes a bi-directional LTSM to learn features and outputs the torsion angles of amino-acid residuals sequentially to build the overall backbone. Although the model only uses the input sequence and its PSSM instead of raw MSAs and additional features, it achieves comparable performance to top results in CASP13 with significantly faster inference speeds at just a few millisconds. Extensive work in NLP has shown transformer models are superior than RNN structures in capturing long range dependency however, and it is questionable whether iterations upon RNN architectures will catch-up to Alphafold2's performance.  

\textbf{GANs - } Although GANs have not been used competitively for structure prediction in CASP or CAMEO competitions, some works have used them creatively for structure reconstruction. \cite{anand2019fully} used GANs to show that it is possible to reconstruct masked components in contact distance maps by treating it as an image in-painting problem. The in-painted distance maps can be used to accurately reconstruct 3D structure. 
% What's more, they discover that by interpolating between the latent spaces of different structures from their generators, they are able to discover valid backbone structures with implications for protein design. 
Such methods have also been shown useful for a special class of structure prediction known as loop modelling, where segments of known structures from irregular, sometimes flexible, loops that are hard to determine even experimentally \cite{ruffolo2020geometric}. \cite{li2017protein} similarly treats loop prediction as an image in-painting problem by using GANs to estimate the inter-residue distance map for loop segments. 

\subsection{Summary}
% https://moalquraishi.wordpress.com/2020/12/08/alphafold2-casp14-it-feels-like-ones-child-has-left-home/amp/#s4.1

The extra-ordinary performance of Alphafold2 is a great leap forward in our understanding of structure and vastly outstrips existing approaches.
% has made a profound impact in bioinformatics. The average error rate of their model predictions are comparable to the error rate from crystallography experimentation, thus significantly reducing room for further improvements on the benchmark task. 
Despite this, there are still important problems in structure prediction that remain to be addressed. Essential life functions tend to be carried out through multi-protein complexes, where interactions between multiple structures drive vital processes. Individual protein structures within these complexes remain challenging to model and are outside the scope of CASP. Also, the current accuracy of Alphafold2 in RMSD terms of 1.6\r{A}, which is highly impressive but still too large for direct application in drug design, where confidence of individual atoms needs to be within 0.3\r{A}. Sub-classes of structure prediction problems, such as loop modelling, which are not as well-defined, also have room for improvement.  Thus, there remains open topics for research in the field of structural bioinformatics for proteins.  %https://occamstypewriter.org/scurry/2020/12/02/no-deepmind-has-not-solved-protein-folding/
% https://www.nature.com/articles/d41586-020-03348-4

% Computation requirements remain a significant barrier, as Alphafold2 was trained with 128 TPUv3 cores, equivalent to 100-200 GPUs, which is out of reach for most research teams. Community efforts such as RoseTTAFold have shown promising attempts at achieving similar performance with limited resources, although there is still significant room for efficiency improvements in methodology. 

% Structural prediction from sequence is the most well defined category of problems within protein bioinformatics. 

\section{Functional prediction}

% correlate proteins based on function by looking at structure. 

In this section, we look at the deep learning methodologies used for functional prediction. Unlike structure prediction, where the task is specific, well-defined, and bench-marked by regular competitions, the problem of functional prediction is much wider in scope, often with varied targets and objectives, such as enzyme/non-enzyme, solubility,  toxicity, and Gene Ontology (GO) classification. Functional prediction can also take place on both a global scale, such as protein classification, or local scale, such as binding site and phosphotylation site prediciton within a protein structure. We also look at the problem of functional prediction from two input domains: sequence and structure.

\subsection{Function from sequence}

 Function from sequence involves functional prediction by directly using the amino-acid sequence as input, thus operating on the 1-Dimensional input space. Functional prediction can be approached by directly training a classifier based on labels for functionality, which is a supervised learning approach. Labels for desired functionality can be sparse however. For example, the GO database \cite{ashburner2000gene} only contains labels for <1\% of if the proteins in UniprotKB, which number above 70 million. Advances in deep learning within other domains, in particular language modelling for NLP, have shown that unsupervised or semi-supervised learning approaches using vast unlabelled data can significantly boost performance. Attempts have also been made to carry these idea to the task of functional prediction for proteins. We investigate both supervised unsupervised approaches in the following sections. 

\subsubsection{Fully supervised approaches}

\textbf{GO functional prediction -} One of the more well-defined challenges in functional prediction is prediction of Gene Ontology (GO) function labels. The GO database contains over 40,000 functional classes, and each protein is labelled with the functions it must perform in order to function \cite{ashburner2000gene}. The labels only represent less than 0.1\% of all the known proteins in the UniProKB database however \cite{uniprot2019uniprot}. The community challenge, CAFA (Critical Assessment of protein Function Annotation algorithms), is held in 2-4 year intervals, with the results of CAFA4 still under assessment \cite{zhou2019cafa}. The official winner for CAFA3, announced in 2019, was GOLabeler \cite{you2018golabeler}, which uses features extracted from multiple statistical models including BLAST-KNN \cite{radivojac2013large} and ProFET \cite{10.1093/bioinformatics/btv345}, and ensembles them in a ranking model to find top K predictions. Further iterations by the authors included NetGO \cite{10.1093/nar/gkz388}, which used additional information from the STRING database \cite{szklarczyk2016string} to improve performance. 

Deep learning approaches to the problem was first applied in DeepGO \cite{kulmanov2018deepgo}, where the authors used an embedding layer, followed by a single 1D convolution layer for feature extraction. Because of the hierarchical nature for some functional classes in the GO labels, a series of hierarchical binary classifiers were also used for final class prediction. The results achieved surpased GOLabeler and NetGO. DeepGO+ \cite{kulmanov2020deepgoplus} improved results further by simplifying the input and output modules of the architecture and increasing the number of 1D CNN layers. 
% \cite{cao2021tale} TALE: Transformer-based protein function Annotation with joint sequence--Label Embedding, beats, but unsupervised, not by large margin. 

More sophisticated architectures such as RNNs and Transformers have also been used. One of the challenges from dealing with protein sequences is the large variation in sequence length. \cite{ranjan2019deep} uses bidirectional LSTMs to condense input sequences into a fixed-size feature vector, which is then used to classify GO functional labels, although performance is not compared with \cite{kulmanov2018deepgo,kulmanov2020deepgoplus}. TALE uses the Transformer architecture \cite{cao2021tale} to perform GO label classification but with an additional label embedding. In order to capture the hierarchical nature of GO labels, labels are represented as a directed acyclic graph and embedded as an additional feature input. They were able to achieve state-of-the-art results for 2 out of 3 problem subclasses and competitive results with NetGO and GOLabeler for the remainder. Notably, pre-training the Transformer through unsupervised training did not lead to performance improvements.

\textbf{Site-specific prediction tasks - }Supervised deep learning has also been used effectively on local classification tasks. On such problem is prediction of protein phosphorylation sites, which are locations on a protein structure that are modified by covalent bonding of a phosphate group. This process takes place as part of the regulation process for protein activity. Deep-phos \cite{luo2019deepphos} uses segments of the protein sequence as input, and a novel CNN block, which they term DC-CNN, for feature extraction. Each DC-CNN is based on 1D convolutional operations with connections between earlier layers. Outputs from 3 DC-CNN blocks with different window sizes are then aggregated for feature extraction, which are then classified as phosphorylation or non-phosphorylation sites. 
\cite{ranjan2019deep} also performs phosphorylation site classification with their MUscADEL model. MUscADEL uses a standard Bidirectional LSTM followed by a fully connected layer to perform site classification for different locations of the protein sequence. With this architecture, they were able to improve upon state-of-the-art performance of more traditional statistical methods.

\subsubsection{Unsupervised / Semi-supervised approaches}

Supervised learning limits the usable dataset to only those with annotated labels. This ignores the vast amount of unlabelled data available for protein sequences. Similar to language models in NLP, where valid sentences must follow generally accepted grammar rules and word relations, naturally occurring protein sequences must obey the natural constraints that allow them to fold into functional structures. Consequently, unsupervised and semi-supervised techniques have been used to try and capture this underlying information. Most approaches attempt to demonstrate generality of learnt representations however in contrast to benchmarking perfomance on a specific task. 

\textbf{Learning latent representations - } The authors in \cite{das2018pepcvae} use an VAE model with RNN encoder and decoders to learn latent representations of antimicrobial peptide (AMP) sequences. The model is first trained with reconstruction loss using 1.7 million unlabelled sequences from the Uniprot-unlab-1.7M database and 15K sequences from AMP-lab-15K in a unsupervised fashion. The RNN encoder is able to learn a latent representation for input sequences through this process. The model is then fine-tuned using the labelled AMP-lab-15K  data by training two classifiers, one using the latent vector as input and one using the decoder output as input, through multi-tasking. The use of unlabelled data significantly increases the training dataset during unsupervised learning, allowing more relevant latent representations to be learnt. 

\textbf{Methods from NLP - } Doc2Vec is a two layer nerual network used frequently in NLP tasks for learning word embeddings. They can be trained with the continuous bag-of-words task, where the network predicts a central word based on surrounding context. \cite{yang2018learned} trains a Doc2Vec model using K-mers (short sequences of length k) from protein sequences to learn sequence embeddings. The embedding outputs from the trained model is then used to perform a set of 4 different tasks. They show that their results outperform the models ProFET \cite{10.1093/bioinformatics/btv345}, which are based on manually extracted features, and are competitive with using one-hot encoded inputs, which have much higher dimensionality. \cite{heinzinger2019modeling} uses transfer learning by fine-tuning a pre-trained ELMo (Embeddings from Language Models) model as the embedding model. The model was then trained for protein-level (water-solubility classification) and residue-level classification (secondary-structure classification) tasks. They were able to achieve competitive results but unable to surpass state-of-the-art results however, which were based on supervised deep learning \cite{almagro2017deeploc, klausen2019netsurfp}. 

The UniRep model in \cite{alley2019unified} was one of the first large scale unsupervised representation learning studies for protein sequences. Unlike \cite{heinzinger2019modeling} which used an ELMo model pre-trained for NLP tasks, Alley et al. trained an LSTM from scratch using unsuprevised learning with around 24 million protein sequences from  UniRef50 \cite{suzek2015uniref}. The network was trained to predict the next amino acid in a given sequence, learning a feature representation of size 1,900. The learnt feature embeddings were found to form meaningful clusters that correlated with their source organisms, and could be used to predict structural classification of proteins (SCOP) classes by training lightweight classifiers on top of the feature extractor. Further experiments also showed that a sparse linear model trained on top of the features could be used to predict a wide variety of quantitative measures such as protein stability, functional ability of mutant sequences, and fluorescense level of mutant proteins. 

\textbf{Self-attention with Transformers - }  \cite{rives2021biological} trains a Transformer using random residue masking and recovery to learn unsupervised representations. Similar to \cite{alley2019unified}, they observe learnt features were able to form semantically relevant clusters. When fine tuned with a classifier, the transformer model was able to better predict contact points, secondary structures, and mutational effects then traditional LSTM models . \cite{nambiar2020transforming} also performs a similar study and conducts additional experiments, showing that the features can accurately predict protein families as well as protein-protein interactions. \cite{vig2020bertology} performs an in depth analysis of attention weights on three separate Transformer models, Bert, ALBERT, XLNet, trained with protein sequence data. Attention weights are examined to see how they respond at contact point sites in structural folds, binding sites in PPI, and PTM sites for protein regulation through various visualization tools. 

% side question for myself, what about using contact maps? Ans: actually contact maps sort of like graphs (vertices and edges)

\subsection{Function from structure}

Functional properties of proteins are largely determined by their structure in their 3D space. Even though structure itself is also largely determined by sequence, performing functional prediction based on structural inputs allows the model to learn from inter-residue relations in their folded state. This requires modelling approaches outside of 1-dimensional sequence-based input space however. Geometric Deep Learning methodologies are used instead, which in itself is also a relatively nascent field. Here we review techniques using two classes of geometric deep learning: deep learning on graph domains and manifold domains.

\subsubsection{Structural representation using graphs}

\textbf{Protein property classification - } The equivalent of convolution, pooling, and attention mechanisms in the Euclidian plane that form the main building blocks of common architectures are not immediately obvious in the graphical domain. One of the earliest pre-curser works to GNNs is \cite{duvenaud2015convolutional}, which explores differentiable convolution operators on molecular graph data. Prior to their work, there was no defined approach to treat graphical inputs of varying size. Molecular fingerprints based on hash functions for substructure encoding were used for sate-of-the-art feature extraction \cite{rogers2010extended}. \cite{duvenaud2015convolutional} replaced the hash function with a differentiable linear operator to combine neighbourhood features and the tanh activation function. This operation was then extended to multiple layers, extending the receptive radius, before aggregation with average pooling. By examining feature activations, they were able to confirm that the network was able to identify useful features such as carcinogenic sub-structures when trained to classify for toxicity.  \cite{coley2017convolutional} also performed a similar study by using dense convolutions with neighbouring atoms within the molecule to perform feature transformation. Different from \cite{duvenaud2015convolutional} which only considered the atomic element as the only input, they also included standard atomic properties such as particle charge and polar surface area. They showed that their learned features could outperform \cite{duvenaud2015convolutional} in tasks such as solubility and melting point prediction. %quantitative structure-activity or property–structure relationships (QSAR or QSPR, often used interchangeably)

Subsequent advancements in Graphical Neural Networks frequently used graphical data from molecules as generic benchmark tasks. \cite{niepert2016learning}, the first paper to formalize the framework for GNNs, used the PROTEINS and D\&D dataset to train a classifier for enzymes and non-enyzymes, achieving ~75\% accuracy on both. Vertices of the graph represented backbone atoms, and edges represented close proximity in 3D space. \cite{ying2018hierarchical} improved upon the GNN network by introducing differentiable pooling operations that can be used to learn hierarchical representations of subgraphs. Performance was similarly benchmarked on PROTEINS, D\&D, and an additional ENZYMES dataset and demonstrated to improve performance over \cite{niepert2016learning}. \cite{lee2019self} further boosted performance by introducing self attention mechanisms to the graph structure by using convolution layers to learn an attention mask on nodes. 

\textbf{Protein-protein Interaction (PPI) - } GNNs have also been used for more function specific tasks, such as ligand-receptor prediction. \cite{fout2017protein} demonstrated that GNNs can be used to predict ligand-receptor relationships in protein molecules. The ligand and the protein are both represented in graphical format and passed through a GNN for feature extraction. The features are then concatenated and passed through a fully-connected layer for classification, achieving accuracy of up to 90\%. \cite{pittala2020learning} also uses GCNs for PPI prediction, but specifically for antibody and antigens. Because of limited data on antibody-antigen pairing, transfer learning is used by first training on general protein-protein interactions. An attention layer is also used to aggregate features extracted from antibody and antigen graphs.

\cite{torng2019graph} also uses GNNs to look at protein-ligand interactions, but by specifically looking at binding-site pockets, defined as the residues within 6\r{A} of the bound ligand molecule. The DrugBank database was first use to find protein structures from the PDB that bound to drug ligands. The pocket of the 3D structure was found and converted to a graph by denoting atoms as vertices and forming edges between atoms within 7\r{A} of each other. Two separate GNNs were trained, one for the protein pocket and one for the target ligand. Features were then combined through a fully connected layer before softmax was used for classification. To ensure relevant features are learned, the pocket GNN encoder was initialized with weights of a graph auto-encoder trained unsupervised using protein pocket graphs. Learnt pocket features were found to cluster according to their SCOP. Feature importance visualizations also showed which parts of ligands and proteins contribute most to binding prediction.

\begin{figure*}%
\centering
\includegraphics[width=0.75 \columnwidth]{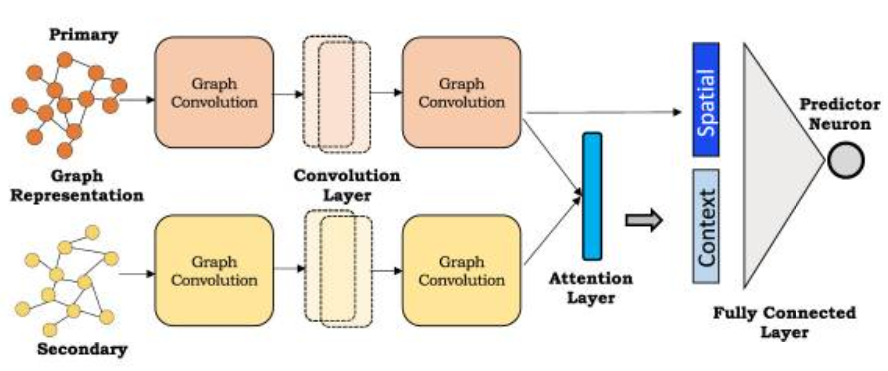}%
\caption{The flow diagram shows an example of how GNNs can be used for PPI or protein-ligand classification. Features from the two molecules can be extracted using a singe GNN or two separate GNNs depending on the application. The extracted features can then be used for downstream classification. Figure is taken from \cite{pittala2020learning}}
\label{gnn_example_drug}
\end{figure*}

\subsubsection{Structure as manifolds}

\textbf{Site-specific classification - } One of the limitations of graph-based approaches recognized early on was that graphs do not fully capture 3D information in its inputs \cite{coley2017convolutional, kearnes2016molecular}. Although 2D atomic distance is reflected somewhat in their edges, a lot of 3D information remains unused. 3D structure information is particularly important for functional prediction however as they are the major determinant of functionality. Geometric deep learning approaches can be used to directly perform learning on surfaces by moving from Euclidean space to Riemannian manifolds, which requires the use of a new set of feature operators \cite{monti2017geometric, masci2015geodesic, boscaini2016learning}. This subclass of geometric deep learning in itself is still relatively nascent however.

\begin{figure*}%
\centering
\includegraphics[width=0.85 \columnwidth]{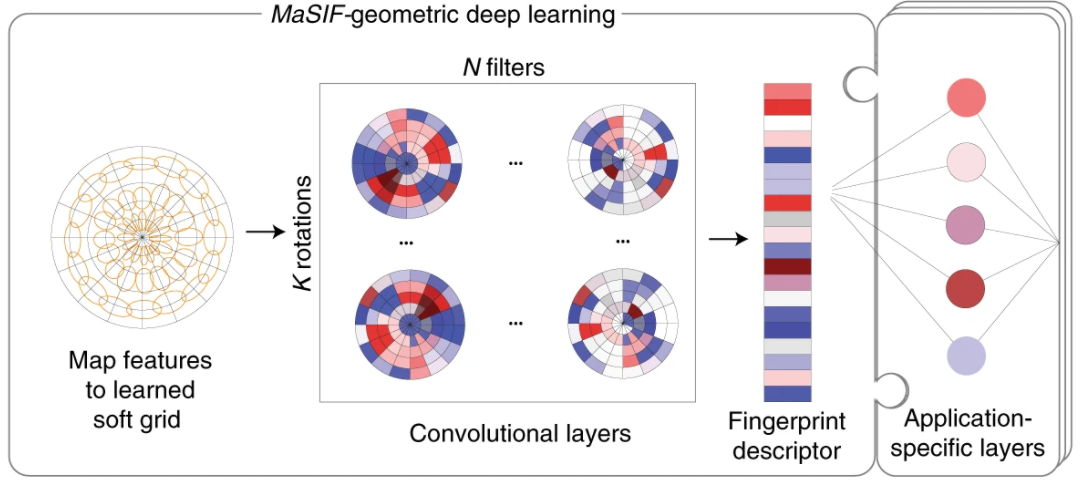}%
\caption{Geodesic convolution applies filters on geodesic patches of a target surface. The features after convolving with the filters form a learnable feature descriptor that can be used for downstream applicatons. Figure is taken from \cite{gainza2020deciphering}}
\label{masif_example}
\end{figure*}

MASIF \cite{gainza2020deciphering} is one of the first works that attempts to perform convolution on surfaces of proteins for deep learning. In Euclidean space, the shortest distance between two points is a line, whereas for manifolds it is a geodesic, a curvature along the surface. The convolutional filters used in CNNs are replaced by geodesic convolutions, which are convolutions within geodesic patches, performed in a sliding window fashion along surfaces. In order to ensure rotational invariance, K rotations are performed on the input patch before convolving with the geodesic filters and the max values are taken. The input features consist of two precomputed geometric features that encode shape and curvature, and three chemical properties of the atom. These inputs are convolved with other atoms within the geodesic patch using the learnt filters to output a series of features at regular point along the surface. These features are then used to perform protein pocket-ligand classification, interface site prediction, and PPI search, surpassing state-of-the-art methods largely based on fixed descriptors \cite{konc2015probis, ritschel2014kripo, shulman2004recognition, xue2015computational, yin2009fast, duhovny2002efficient, pierce2011accelerating}. A simplified diagram of this process is shown in Figure \ref{masif_example}.

%https://www.dropbox.com/s/lug94g9mn02aleq/AMMI%20Lecture%206%20-%20Graphs%20and%20Sets%20II.pdf?dl=0

One of the major limitations of MASIF however is that geodesic convolution requires computationally intensive pre-computation of a surface mesh for surface distance calculation. The geometric features, which are hand-crafted and computed based on the surface mesh, are also non-differentiable. dMASIV \cite{tubiana2021scannet} proposes a new computationally efficient, end-to-end framework, for manifold learning. Instead of representing the input as a pre-computed surface mesh, point clouds of the atoms are used with a sampling scheme to obtain random points and surface normal values along the surface. A novel coordinate transformation function based on surface norms is also applied that allows quasi-geodesic convolution to be performed. Geometric features are replaced with Gaussian curvatures calculated from surface norms. The combined effect is a framework that is more accurate than MASIF and faster by a few orders of magnitudes. 

\subsubsection{Structure from 3D space}

An approach by \cite{tubiana2021scannet} that has yet to be peer reviewewd proposes ScanNet which operates directly in 3D space. Inputs are represented as point clouds, and for each atom, its 16 closest neighbours are used for convolution. A 3D coordinate frame is first formed, orientated based on its covailent bonds with its neighbours, and geometric features are captured through convolving 3D distance from its neighbours with a Gaussian kernel. Chemical features are learnt through a dense layer after product with geometric features. Cross- and self-attention are also applied using a neighborhood attention model. Comparison with MASIF also showed superior results on protein-protein binding site predictions, although inference was dominated by the time taken to perform MSA search as feature inputs, taking one to a few minutes.

\subsection{Summary}

Protein functionality is largely defined by its structure, making structural features a key source of information. Although structure itself is largely determined by a protein's sequence, the challenges in predicting structure from sequence show that relevant features may not be effectively learnt when using sequences directly. Current work on prediction from structure remain relatively limited however since geometric deep learning techniques are still nascent. Functional prediction from sequence may also be improved given the recent breakout performance of Alphafold2 since features relevant to structure prediction should intuitively encompass information on functionality. Whilst unsupervised / semi-supervised methods have shown that useful representations can be learnt using such methods, they have yet to be used effectively to boost functional classification tasks \cite{cao2021tale, heinzinger2019modeling}. This could be a promising source for future improvements given the amount of structured but unlabelled data that is available.

\section{Protein Design}

Protein design is the task of finding amino-acid sequences that fold into stable structures and perform some target function. This is extremely challenging as it can be seen as the reverse of structural and functional prediction from sequences, which themselves are not well understood. Measuring accuracy is also difficult as true performance can only be determined through experimentation. Designed proteins that have passed \textit{in silico} validation, \textit{i.e.} through computational methods, are frequently found to perform poorly when experimented \textit{in vitro}, \textit{i.e.} after synthesis using recombinant DNA technology, due to unstable folds. In this section, we divide protein design into two categories: Sequence from function and sequence from structure.

\subsection{Sequence from function}

\textbf{Generating from latent representations - } One approach for sequence generation has been to learn a representation space for valid protein sequences. New sequences are then generated by random sampling from this latent space. \cite{das2018pepcvae} takes this approach by training an Variational Autoencoder with RNN encoder and decoder structures to first learn latent representations using unlabelled protein sequences. The latent representations were then fine tuned with labelled AMP sequences and used for controlled generation of AMP protein sequences. Sequences were not validated \textit{in vitro} however but suggested as part of future work. Instead, 11 out of 5,000 high probability AMP sequences were selected from the outputs and folded using PEP-FOLD3 \cite{lamiable2016pep}. Out of these, 9 sequences exhibited a helix, which is a typical structure observed in AMPs, thus qualitatively confirming the feasibility of this model. 

\textbf{Generating from auto-regressive models - } Generation can also be achieved through an auto-regressive framework. \cite{riesselman2019accelerating} trains a auto-regressive model based on 1D CNNs to learn the underlying sequence grammar for anti-bodies. The model is fit to the \textit{naive llama immune repertoire library of antibodies} \cite{mccoy2014molecular}, and then used to generate additional anti-bodies sequences. Initial experiments were said to demonstrate valid functionality, although details were not provided. The study also demonstrated an additional use case for the trained model, which is to estimate the likelihood of a mutant sequence being observed, which is often used as an indicator for mutational effect \cite{hopf2017mutation,riesselman2018deep}. The residue output probabilities were used to calculate this measure, and found to have high correlation with values based on realistic observations.

\cite{muller2018recurrent} perform a similar work to \cite{das2018pepcvae} on AMPs using RNNs but with the next in sequence training scheme. Given a starting token, the trained network could then infer a chain of the most likely residues to appear next in sequence. Variability can also be obtained by adding noise within the inferred sequences. \textit{In vitro} exerimentation was not used to validate the results however. Instead, a separate state-of-the-art AMP classifier, CAMP \cite{waghu2014camp}, was used, showing that 83\% of the generated sequences were likely to be AMPs. 

This same technique was used later by the same team in \cite{grisoni2018designing} for anticancer peptide (ACP) generation with minor adjustments. In this study however, the generated sequences were synthesized and validated to be effective through \textit{in vitro} experiments. Due to the relative scarcity of known ACPs, the RNN network was first pre-trained on 10,000 amphipathic peptides before fine-tuned with 26 known ACPs. Generated sequences were than ranked and screened extensively based on confidence scores and feature similarities with known ACPs. 12 of the most promising sequences were synthesized, of which 10 showed activity against cancer cells and 6 effectively killed target cancer cells. These results clearly demonstrated the feasibility of generative methods for realistic protein design. 

\textbf{GANs - } \cite{repecka2021expanding} uses a GAN network trained on malate dehydrogenase (MDH) enzymes and examined the ability of the network to learn latent representations of different properties. An desirable attribute of GANs over other techniques such as autoregressive models is that a latent vector can be used to interporlate between desired properties. Latent vectors were found to be correlated with a number of the protein characteristics such as hydrophobic, charge, and sequence length properties, which allowed interpolation between them to generate desired features. Extensive experiments were performed to determine if the designed proteins could fold into the desired structure with functional catalystic sites. 60 sequences were selected, of which 55 were successfully translated into a DNA expression vector. The protein was produced using re-combinant DNA techniques through E-coli, of which 19 of which were successful and 13 of which expressed the desired enzymic activity, demonstrating the feasibility of such methods. It was noted however that generated sequences closer to the distribution of the training set showed higher probabilities of success. 

\textbf{GANs with structure oracle - } \cite{gupta2019feedback} proposes the Feedback Gan (FBGAN) to directly generate gene encodings for desired protein functionality. The GAN is trained to generate genes for AMP expression with an additional constraint on existence of an $\alpha$-helices in the secondary structure, since these are frequently observed in AMPs. In addition to a generator and descriminator, a functional analyser, PSIPRED \cite{buchan2013scalable}, was used to output the likelihood of a $\alpha$-helix secondary structure based on the sequence. Outputs from the generator were first filtered by the PSIPRED to retain sequences above a cut-off score. These are then fed into the discriminator for training. Their framework allows for different non-differentiable black-box predictors to be used as the functional analyzer to encourage different properties. To evaluate the merits of FBGAN over traditional GANs, they used an independent AMP prediction server, CAMP, to show that 40\% of FBGANs generated outputs that were predicted to be AMPs compared to 32\% without the functional analyzer. They noted difficulties in generating longer sequences compared to shorter sequences using the GAN framework however. 

\textbf{Directed evolution with a structure oracle - } Another technique for sequence discovery is directed evolution, which is inspired by nature. The evolutionary search methodology typically involves local mutations from at least one functional parent that explores a locally smooth sequence-function space, and recombinations between different sequences to allow for jumps in sequence space \cite{romero2009exploring, drummond2005conservative}. Directed evolution involves an additonal screening process to discard non-functional sequence off-springs. Machine learning has been shown to effectively augment this progress by utilizing information from non-functional sequence for next generation evolution as well. \cite{bedbrook2019machine} uses a partial least squares model to estimate the effect of different individual mutations on the functionality of the protein after screening. Positive mutations were then reused on the parent to obtain the next generation. \cite{romero2013navigating} uses a Gaussian process (GP) model with baysian learning to first estimate thermostability of a target protein. The model was then used to reduce uncertainty in the next generation by choosing sequences estimated to have the desired thermostability but maximum variantion with existing sequences. \cite{biswas2021low} also uses this idea but trains a functional classifier using the UniRep feature representation model. The functional classifier was used to output a fitness function to guide directed evolution. Results were not validated experimentally \textit{in vitro} however.

\textbf{Reinforcement learning with structural feedback - } The concept of directed evolution is also similar to that of reinforcement learning, where optimum series of actions are predicted given a current state. \cite{olivecrona2017molecular} augments an RNN trained through unsupervised learning with a reinforcement learning agent that fine tunes generated outputs based on a scoring function. The scoring function can be a trained functional prediction model or any predictor designed to enforce some desirable property. The model was used to generate new molecules with disearable properties and validated using RDKit \cite{landrum}. %RDKit: open source cheminformatics. Version: 2016-09-3. http://www.rdkit.org/. 
The use of the agent was shown to significantly improve performance of a simple generative model. Similar ideas of augmenting GAN networks through reinforcement learning can be found in \cite{popova2018deep, sanchez2017optimizing, putin2018reinforced}. These concepts are also similar to  \cite{gupta2019feedback} and \cite{biswas2021low} in that an oracle for functional prediction is used to direct generation and exploration. 

% Limited by effective sequence-function predictor which is non-trivial. 

% Descriminator allows us to use sequence to function knowledge to guide training, since it is hard to go in reverse like structure to sequence. 

\subsection{Sequence from structure}

Some studies approach the problem of protein design by fixing a desired target structure and finding a viable sequence with the targeted fold. This helps reduce the scope of the problem since functional prediction from structure is still not fully understood. This process involves first fixing a target $C_\alpha$ backbone before determining which residue side-chains can be validly packed to the structure. Similar to sequence prediction from function, evaluation of generated sequences remains a challenge however, since true accuracy can only be determined through physical experimentation. 

\textbf{Early works - } \cite{kuhlman2000native} tries to answer the question of how large is the space of valid sequence chains given a target backbone structure. For a given structural backbone taken from some native protein, Metropolis Monte Carlo search was used from multiple seeds to find the low entropy structures given a latent energy function. They found that for these simulated structures, 51\% of the residues in the core structure are identical to the native protein and 27\% of all residues were identical. This suggested that the space of valid sequences are limited to some extent, with constraints imposed mainly due to stability requirements in the core structure. Although the discovered sequences were not validated through synthesis, they note that experimental results in \cite{maxwell1998mutagenesis}, which conclude that sequence conservation is the best predictor of protein stability, reach a similar conclusion. 

Based on this observation, early studies attempt to use sequence recovery rate of the native protein as a metric for evaluating protein design tasks. SPIN \cite{li2014direct} uses neural networks to identify protein sequences based on overall structural entropy. State-of-the-art approaches for protein design at the time involved either mutating from randomly initialized profiles to find a minimum energy configuration, or assembling structurally similar fragments based on template libraries. The latter methodology is more computationally efficient but less effective at optimizing the structural energy profile. SPIN uses a two layer neural input to predict the output residue given a position. The inputs include the backbone torsion angles, a fragment based sequence alignment profile, and a latent energy value based on the overall structure. Sequence recovery was used as the metric to evaluate suitability of the generated sequence, with SPIN improving the metric from 24\% to 30\% for fragment based techniques. SPIN2 \cite{o2018spin2} was proposed in a later work, improving upon the original model by increasing the number of neural network layers and including additional features. Recovery rates improved to above 34\%. It remains unclear how suitable this indicator is however, given that vastly different sequences are known to conform to the same structure, and that a reconstruction score near 100\% defeats the purpose of finding new varied sequences.

\textbf{Structure as prior - } \cite{greener2018design} uses a CVAE to generate proteins with metal binding properties (metalloproteins). Instead of mutating from a randomly initialized sequence. Structural constructs were enforced through topology based grammar strings. Generated sequences were further sampled in an iterative procedure to determine high confidence sequences. The most promising structures were first selected based on SCWRL4 \cite{krivov2009improved} and Rosetta \cite{rohl2004protein}. The latent representations for these structures were then found using the trained encoder and further sampling was conducted from the distribution of these latent vectors. Validation was done \textit{in silico} through molecular dynamics simulation by checking that the sequences formed stable folds. 

\textbf{Structure prediction as an oracle - } Similar to the concept of FBGANs in \cite{gupta2019feedback}, \cite{karimi2020novo} uses a fast structure-prediction oracle to guide a conditional Wasserstein GAN network to a desired target structured. The GAN network was first pre-trained with unlabeled sequences from UniRef50 by fixing the conditional input to its mean representation value. The network was then fine-tuned with its structural latent input using data from  SCOPe v.2.07 \cite{chandonia2019scope}. An oracle model was trained to predict the probability of the 1,215 folds in the SCOPe database, which is used for feedback when training the generator through incorporation into the loss function. Although \textit{in vitro} evaluation of generated structures was not conducted, the authors use a number of state-of-the-art structure and structural stability predictors to show that the sequences are valid.

\begin{figure*}%
\centering
\includegraphics[width=0.75 \columnwidth]{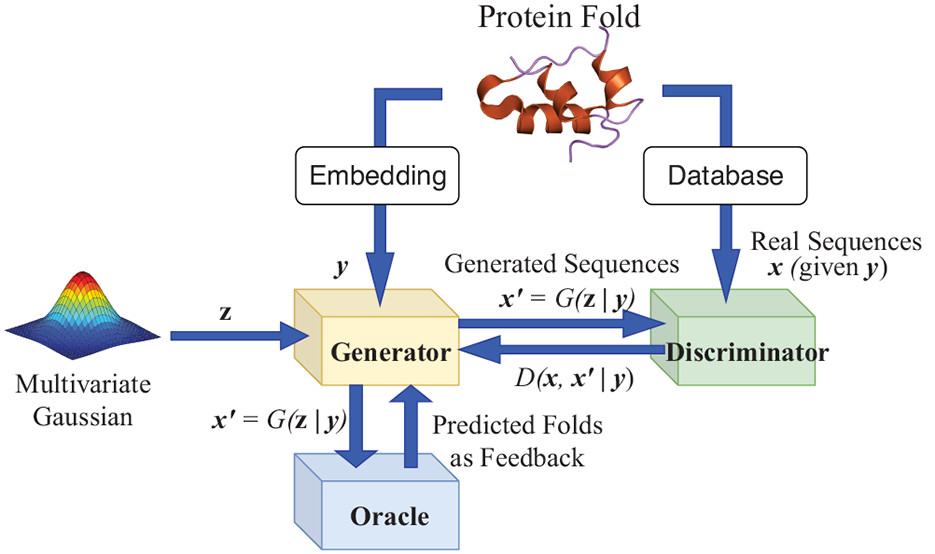}%
\caption{The network diagram above shows an example where a structure prediction model can be used as a oracle for feedback in GAN networks. Figure taken from \cite{karimi2020novo} }
\label{log_graph}
\end{figure*}

\textbf{Encoding structure with GNNs - }\cite{ingraham2019generative} proposes an innovative approach by introducing a Structured Transformer architecture. One of the characteristics of protein structure is that long-range dependencies in sequences are generally result in short-range distances in 3D space. Based on this observation, an encoder-decoder architecture is used, with GNN used for the encoder to capture structural information and a Transformer for the decoder to output the generated sequence. In order to accurately capture the characteristics of the atoms in 3D space, additional spatial and positional encodings are included as part of the graph features. Self-attention mechanisms are also applied. Generated sequences were not validated through \textit{in vitro} experiments, but by a combination of using reconstruction scores and calculating likelihood probabilities for naturally existing sequences using the trained model.  

\textbf{Protein to drug translation - }\cite{grechishnikova2021transformer} uses the Transformer architecture but applies a different take on the problem of molecular design. Instead of generating a target protein sequence, they show that it is possible to design molecules that bind some target protein given its sequence, with direct implications for drug design. Using a Transformer based encoder-decoder structure, they feed in the protein sequence as the input and obtain generated SMILE molecule strings as output, effectively treating the problem as a translation task. Binding data from BindingDB was used to train the model. RDKit \cite{landrum} was used to evaluate the chemical validity of the drug and SMINA \cite{koes2013lessons} was used to check for docking of the generated molecules to target binding cites.

\textbf{Design as a constraint satisfaction problem (CSP) - } %In the problem of structure prediction from sequence, methodologies like \cite{wang2017accurate} approach the task as a pixel classification problem. 
\cite{strokach2020fast} compares structure prediction to a CSP such as solving a Suduku problem. They first identify an optimum architecture through training on generated Sudoku problems. They then apply the same architecture to the problem of sequence search given a target structure. The target structure is represented through a distance map, which also serves as the edges of the graph. Node features on the 2D distance map represent the target amino acid. The network is initially trained to complete sequences chains using training data with masked out node values. To generate entirely new folds, the network is used to infer the node values based on 4 graphs with node values entirely masked out. They find that the generated sequences share approximately 30\% residue identity in their validation set. \textit{De novo} design of entirely masked out sequences were then tested \textit{in silico} to demonstrate nearly identical folds to the target. Further \textit{in vitro} testing based on synthesis from E-coli showed that the folded structures correspond to expected values.

\begin{figure*}%
\centering
\includegraphics[width=0.9 \columnwidth]{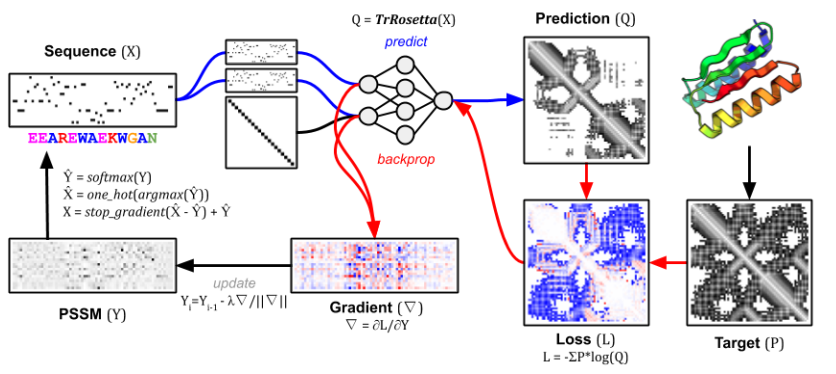}%
\caption{ A trained network can be used through gradient ascent to perform back-propagation directly on inputs. This allows inputs that correspond to some desired output target to be generated directly. Figure taken from \cite{anishchenko2020novo} }
\label{grad_ascent}
\end{figure*}

\textbf{Reversing structure prediction models - }Most of the above approaches use structural information either as a pre-determined constraint or a feedback oracle. It has long been recognized however that sequence design from structure can be seen as the inverse problem of structure prediction from sequence. Recent advancements in state-of-the-art results for structure prediction and deep learning methodologies have allowed this approach to be utilized in protein design. Information and features learnt by a high-performance structure prediction network can be reversed through freezing model weights and back-propagating on a randomly initialized input sequence directly. This process is known as gradient ascent, or alternatively deep hallucination, and leads to modified inputs that are consistent with some target output. An example of this is shown in Figure \ref{grad_ascent}. \cite{anishchenko2020novo} demonstrates the feasibility of this idea through deep hallucination using the trRosetta \cite{yang2020improved} network. A random sequence was first used to find the corresponding noise distribution of contact distance maps. Gradient ascent was then used to generate new sequences with the objective of maximizing KL divergence of the distance maps with the noise distribution to generate highly varied structures. To demonstrate that the hallucinated sequences are valid structures, 129 of them were synthesized through DNA recombination in E coli, out of which 27 successfully folded into stable structures consistent with the trRosetta network predictions.

Whereas \cite{anishchenko2020novo} uses hallucination to demonstrate successful folding of various generated sequences with varying structures, \cite{norn2020protein} applies a similar concept on the reversed trRosetta to find the highest probability sequence of a given target structure. Using Bayesian probability, we see that:

\begin{equation}
    P(sequence|structure) = \frac{P(structure|sequence)\times P(sequence)}{P(structure)} 
\end{equation}

where $P(structure)$ is constant for a desired target. $P(structure |sequence)$ can be directly estimated using the orientation distributions outputted by the trRosetta model, whilst $P(sequence)$ can be estimated based on KL distance between observed residue pair frequencies in PDB and the generated sequence. Thus, $P(structure|sequence)\times P(sequence)$ can be directly back-propagated using gradient descent onto a softmax representation of the sequence. For the entire process, a random sequence is first initialized, from which back-propagation is applied to generate new sequences that have contact maps closer to the desired target. Out of these, the highest probability sequence is used to initialize the next iteration. The procedure was found to perform better than prior state-of-the-art methodologies such as Rosetta by identifying sequences with lower latent energy. 

\cite{tischer2020design} also examines the probability of protein design with targeted binding sites through hallucinating the trRosetta model. Proteins sometimes perform regulation functions by scaffolding the binding site of other target proteins, but there is currently no computationally efficient method for sampling known protein structures to find effective scaffolding structures. Instead of constraining the entire protein structure, the goal is to generate valid proteins with specifically defined functional sites to perform scaffolding. A structural motif representing the active site is first obtained by determining the residual sequence of a known binder that is within 5\r{A} of the binding site. This residual sequence is than split into chunks and fixed as part of the generated sequence. Remaining sections, including gaps between the residual chunks, are hallucinated following the methodology in \cite{anishchenko2020novo}. In addition to KL divergence loss, a structural motif loss function is introduced to ensure that the binding site structure is present in the final fold. Visualizations of the final fold show chemically feasible proteins with scaffolding properties. Although \textit{in vitro} validation is not performed, a variety of simulation and evaluation methods are used to show that the sequences form valid folds with high confidence.

\subsection{Challenges in protein design and the impact of deep learning}

Protein design is an important component of tasks such as bioengineering and drug design, but remains highly challenging. Different sequences may fold into the same target structure, making it a ill-poised problem, and experimental validation \textit{in vitro} is required to ensure generated sequences can maintain stable folds in natural environments, which limits the ability to iterate rapidly. Unlike structural and functional prediction problems that have standardized evaluation metrics, accurate evaluation and comparison between different design methodologies remain difficult. Currently, a large proportion of structures shown to be stable and valid through \textit{in silico} experiments do not fold when synthesized. Thus, attempts to use \textit{in silico} methods, latent energy analysis, or metrics such as recomposition scores for comparisons may not be entirely accurate. In order to truly address this issue, a better understanding of the deficiencies in current \textit{in silico} procedures is required, which also implies a better understanding of folding dynamics and fold stability. Similarly, more focus could be placed on the percentage of \textit{in silico} folds that achieve stable structures when synthesized \textit{in vitro}. 

Recent improvements in structure prediction with deep learning have unlocked new approaches to the problem however, making it the process much simpler computationally. Oracle based feedback in generation and mutational approaches help direct the sequence search to samples with targeted properties. Prediction models encompassing structural and functional knowledge can therefore be used directly as part of generator frameworks. For some structure prediction models, the relatively long inference times make this approach difficult. It has been shown however that the learnt features an correlations in these models can be directly reversed to find sequences consistent with target structures. Outputs can be targeted to specific structures \cite{anishchenko2020novo} or even to perform specific functions \cite{tischer2020design}. These methodologies have only been made available recently through the increased performance of learning based architectures.

Advancements in structural prediction, most notably from Alphafold2, have yet to make its impact in protein design. Given the ability to treat design directly as the reverse problem for structural and functional prediction, developments in these categories should be closely linked. Proper evaluation methods will need to catch-up first however if future advancements are to be properly measured. 

\subsection{Summary}

Sequence generation from function and from structure can be seen as two different problems but with the same motivation. Sequence from function relies on learning some representation or embedding for a given function and generating new sequences based on these learn learnt embeddings. Sequence prediction from structure aims to simplify the problem by limiting its scope and targeting a single pre-defined structure. Out of the three classes of problems, protein design is the most challenging one, limited partly by extra difficulties in performance evaluation. Nevertheless, existing works have shown that improvements in other problem categories such as structural prediction can directly contribute to better methods and performance. Given the importance of the problem, protein design will remain an interesting problem as the entire field of protein bioinformatics advances.

\section{Discussion}

% We have broadly classified problems in protein bioinformatics into three main categories which we explored in Sections 3-5. 
We give an overview of some common hurdles in protein bioinformatics and highlight additional topics that are important to future developments. 

\textbf{Rapid Experimental Validation -} Data collection and validation are essential parts of the machine learning pipeline. Due to advancements in sequencing technology, databases such as PDB have grown rapidly in size, whilst labels for some applications still remain limited. The ability to rapidly determine protein characteristics through experimentation can greatly enhance the ability to address task specific prediction problems in bioinformatics. Experimentation is particularly important in design problems where \textit{in vitro} validation is required to guarantee results. The equivalent process in the CV and NLP domain is qualitative feedback from human reviewers for generative results, which is considerably simpler and cheaper to conduct at scale. Advancements in laboratory experimentation and computational bioinformatics are therefore mutually beneficial, and greater improvements can be achieved through rapid progress in both areas. 

\textbf{Dynamic Modelling of Proteins -} Standardized challenges and benchmarks such as CASP have played an important role in accelerating our knowledge of the folding process. However, it is important to remember that the challenge is only restricted to a specific formulation of the problem. Current collection of structural information used in CASP and most experiments are based on crystallography, where the protein is transformed into crystal form through vapor diffusion. This is an important detail because cells are not crystalline, and experimental structures obtained through crystallography do not truly reflect structure in its natural environment. The actual folding process itself is also a dynamic process that is not well understood. To give some examples, there are still conflicting view for example on whether the process is multi-pathway or path-dependant \cite{englander2017case}, which has implications on where they stabilize on the energy landscape. Some folding processes also depend on chaperone proteins that enable it to make specific folds. Even after stabilization, some proteins may still maintain flexible backbones which make them especially difficult to capture and model. Studying the full dynamic nature of proteins is highly complex, and there is still room for improvement in this field. 
% Deep learning methodologies have not been applied to tackle this challenging topic yet. 

\textbf{Better use of 3D information -} Structural information is important for many problems, especially those involving functional prediction or protein-protein interaction. Machine learning directly using protein structural information is a relatively nascent field however. Problems in CV and NLP tend to settle on a number of standard architectures based on 1D or 2D input domains, but 3D structures can be represented and studied in many different ways by modelling them as surface manifolds, graphs, or local 3D coordinate frames. As of yet, it is not clear which is the best approach. Geometric deep learning frameworks have attempted to generalize deep learning across different input domains through a common framework, but these efforts have been relatively new. There is still room for innovation in these new tool sets.

\textbf{Biophysical priors - } Unlike CV and NLP problems which involve human interpretation, protein bioinformatics are largely determined by biophysical processes that restrict what is feasible. Techniques such as molecular dynamic simulation attempt \cite{hollingsworth2018molecular} to model these physical processes directly through simulation, although these are usually computationally intensive and require accurate knowledge of the underlying forces. Deep learning takes the opposite approach by approximating the underlying processes based on learning from large datasets, which has shown to be effective, but it is not clear whether the approximations are consistent with the laws of biophysics. This has inspired some methodologies where biophysical interactions are used either for model regulation \cite{raissi2019physics} or as a model prior \cite{lutter2019deep}. Other works attempt to directly model molecular forces through machine learning \cite{noe2020machine}. Effectively integrating biophysical knowledge could help increase the performance of current models and improve model explain-ability, and is an interesting topic for future research.

\textbf{Cross-domain tasks between proteins and drug molecules - } Protein bioinformatics is closely related to drug design, as drugs are molecules designed to have specific effects on target proteins. Currently, studies tend to treat these two inputs separately by either acting solely on protein or molecular inputs. \cite{grechishnikova2021transformer} is an example where proteins and drug molecules are jointly considered through translating between two sequences using a Transformer. This is an interesting approach as it allows a more application-specific task to be learnt directly instead of modelling the two separately. Such an approach may lead to more innovative approaches in drug design as it attempts to learn the relevant task directly in an end-to-end manner.

\section{Conclusion}

Deep learning has been increasingly applied to problems in protein bioinformatics to achieve impressive results. The field is well-suited for deep learning methodologies due to the availability of large, structured datasets and the ability of deep learning to approximate complex interactions such as molecular dynamics. Recent works have also shown that such techniques clearly overtake traditional modelling approaches. In this survey, we broadly categorized problems into structural prediction, functional prediction, and protein design, and reviewed how advancements in deep learning have been applied to achieve promising results in each. Unlike the more mature fields of CV and NLP however, proteins are still a relatively new domain for deep learning and new tools and techniques, both computationally and experimentally, remain to be discovered. On the whole, this is a fast developing area with direct social benefits in terms of its potential applications to disease discovery and treatment. We believe this is an area that will attract an increasing number of researchers and we look forward to future advances.

https://moalquraishi.wordpress.com/2018/12/09/alphafold-casp13-what-just-happened/
https://moalquraishi.wordpress.com/2020/12/08/alphafold2-casp14-it-feels-like-ones-child-has-left-home/
https://www.blopig.com/blog/2021/07/alphafold-2-is-here-whats-behind-the-structure-prediction-miracle/
https://lupoglaz.github.io/OpenFold2/msa.html
https://www.icr.ac.uk/blogs/the-drug-discoverer/page-details/reflecting-on-deepmind-s-alphafold-artificial-intelligence-success-what-s-the-real-significance-for-protein-folding-research-and-drug-discovery
https://fabianfuchsml.github.io/alphafold2/

{\small
\bibliography{refs}
\bibliographystyle{plain}
}

\end{document}